\documentstyle[editedvolume,namedreferences]{crckapbk}

\def \eg {{e.g. \/}}

\def \etal {{\it et al.}}

\def \cf {{\it cf. \/}}

\def\apjref#1;#2;#3;#4;#5 {#1 (#2) {\it #3}, {\bf #4}, #5}
\def\ltsima{$\; \buildrel < \over \sim \;$}
\def\simlt{\lower.5ex\hbox{\ltsima}}
\def\gtsima{$\; \buildrel > \over \sim \;$}
\def\simgt{\lower.5ex\hbox{\gtsima}}

\def\kms{${\rm km s^{-1}$} }
\def\kms2{$\rm km s^{-2}$ }

\begin{opening}
\title{CHEMICAL  EVOLUTION OF THE GALACTIC DISK AND BULGE}
\author{ROSEMARY F.G. WYSE}
\institute{Department of Physics and Astronomy, The Johns Hopkins
University, Baltimore, MD 21218, USA (permanent address)\\ and \\
Center for Particle
Astrophysics, University of California, \\ Berkeley, CA 94720, USA\\
(wyse@wyser.pha.jhu.edu)}

\end{opening}

\begin{document}

\section{Introduction}

The Milky Way Galaxy offers a unique opportunity for
testing theories of galaxy formation and evolution.
The study of the spatial distribution, kinematics and chemical
abundances of stars in the Milky Way Galaxy allows one to address
specific questions pertinent  to this meeting such  as

\begin{enumerate}
\item[(i)] When was the Galaxy assembled?  Is this an ongoing
process? What was the merging history of the Milky Way?
\item[(ii)] When did star formation occur in what is now `The
Milky Way Galaxy'?  Where did the star formation occur then?
What was the stellar Initial Mass Function?
\item[(iii)] How much dissipation of energy was there before and
during the formation of the different stellar components of the
Galaxy?
\item[(iv)] What are the relationships among  the different
stellar components of the Galaxy?
\item[(v)] Was angular momentum conserved during
formation of the disk(s) of the Galaxy?
\item[(vi)] What is the shape of the dark halo?
\item[(vii)] Is there dissipative (disk) dark matter?
\end{enumerate}

Chemical evolution allows one to focus on particular aspects of galaxy
formation, such as the stellar initial mass function (IMF), the star
formation history -- including the homogeneity of the star formation
process -- and the rate of mass assembly to form the `final' galaxy.
Elemental abundance ratios play an important role in reaching fairly
model-independent conclusions from the confrontation of theory with
observations.  The scatter about mean trends also can not be ignored,
as was the common practice in the (recent) past, and understanding the
physics of intrinsic variations in the star formation process will be
a major step forward.

\section{Galaxy Formation}

I shall briefly summarise current ideas of galaxy formation and
evolution to place Galactic work in context (see \eg\ Silk and Wyse,
1993 for a more complete review, and White this volume).  The nature
of dynamically-dominant Dark Matter determines the way in which
structure forms in the Universe.  All density fluctuations are
initially just gas and dark stuff, and the rate at which stars form is
very model-dependent, but feedback from stars plays a major role in
the energy balance.  Hierarchical-clustering theories such as Cold
Dark Matter (CDM; \eg\ Blumenthal \etal, 1984) predict that the first
objects to collapse under self-gravity are a small fraction of the
mass of a typical galaxy, so that galaxies form by merging of these
smaller objects.  The rate of merging and growth of mass of a
protogalaxy can be estimated by studying the dark matter, which is
assumed to be dissipationless and hence to have simpler physics than
the baryonic component (but note that this provides an {\it
over-estimate\/} of the merging rate of the dissipated, baryonic
component e.g. Navarro, Frenk and White 1994).  Lacey and Cole (1993)
have a particularly vivid schematic representation of the merging
process as a tree (their Figure 6), where time increases from top to
bottom, and the width of the branches indicates the mass of a
particular halo associated with galactic substructure. The final
product, the galaxy, is represented by the single trunk at the ground,
and the relative thicknesses of branches meeting reflect the mass
ratios in a merger.  The merging history of a galaxy is then described
by the shape of the tree. The extremes of morphological type may
perhaps be the result of merging histories that are described by the
two types of trees that I at least was taught to draw as a child --
either one main trunk from top to bottom, with many small branches
joining the trunk at all heights, as a fir tree, or a main trunk that
splits into two repeatedly, more like an English oak tree.  The
latter, dominated by `major mergers' or equal mass mergers, may lead
to an elliptical galaxy.  The former, where the merging history is
dominated by `minor mergers', or very unequal mass mergers with a
well-defined central core at all times, may lead to a disk galaxy.
This picture of disk galaxy formation -- building up by accretion of
substructure onto a central core -- provides a synthesis of elements
of the much-discussed and previously apparently mutually-exclusive
`monolithic collapse' paradigm of Eggen, Lynden-Bell and Sandage
(1962) and the `chaotic' halo formation envisaged by Searle and Zinn
(1978).  One of the reasons for the popularity of CDM is its analytic
tractibility; Gunn (1987) presents an elegant exposition of the
formation of the Milky Way.  Constraining the merging history -- and
future -- of the Milky Way is of obvious importance, as is
determination of the star formation history.

Early star formation in the proto-Galaxy is likely to occur not only
in the central core, but also in low-mass substructure - perhaps of the mass
of dwarf galaxies -- in the envelope of the density profile surrounding
the local peak in the (smoothed) density fluctuation field that is
destined to become a galaxy.  The onset of local gravitational
instability and star formation in Galactic substructure will not be
simultaneous but will reflect the natural spread in local density
fluctuation amplitude (e.g. Silk and Wyse, 1993).
The shallow depth of the potential well of these
low-mass systems will most probably mean that their chemical evolution
is truncated, with their first generation of stars capable of
producing drastic effects such as supernova-driven winds (e.g. Dekel
and Silk, 1986).  The stellar halo probably formed from the merger of
substructure like this (e.g. Hartwick, 1976; Searle and Zinn, 1978); the
ejected gas will likely flow towards the central regions, based on
angular momentum arguments (Carney, 1990; Wyse and Gilmore, 1992) and
also from elemental abundance considerations (discussed below).
Further build-up of the `bulge' results from dynamical friction
operating on the stellar components of accreted/merged systems and by
gas inflow driven by gravitational torques during the merging
process.  Dissipation, and settling to the symmetry plane, of the gas
which was not intimately involved in earlier star formation provides a
disk. Again, the angular momentum of the disk at the solar
neigbourhood is substantially higher than that of the stellar halo,
and the half-mass radius of the stellar halo is significantly less
than that of the disk, arguing against significant pre-enrichment of
proto-disk gas by mixing with the products of stellar evolution in the
halo; this would suggest that the oldest disk stars -- most probably
now in the thick disk -- should be metal-poor.

Early theories of disk-galaxy formation within dynamically-dominant
dark-matter haloes assumed a fixed angular-momentum vector and
detailed angular-momentum conservation (Fall and Efstathiou, 1980).
More recent calculations have addressed the effects, within tidal-torque
theory for the generation of angular momentum, of the inherent
run of density fluctuations about a peak (e.g Ryden, 1988; Quinn and
Binney, 1992). The spins of shells of material located at distances
$r_i$ from the peak's location tend to be anti-correlated for $r_1/r_2
\simgt 2$ (Quinn and Binney, 1992).  Further, gas viscosity and
dynamical friction will cause angular momentum re-distribution
(e.g. Zurek, Quinn and Salmon, 1990; Katz and Gunn, 1991), augmented by torques
due to non-axisymmetric perturbations such as a bar (e.g. Blitz and
Spergel, 1991).  The net effect is that not only will infall to the
proto-disk be lumpy, due to the underlying spectrum of density
fluctuations, but its angular momentum vector will be time-dependent,
with resulting complicated final equilibrium.  The disk can certainly
{\it not\/} be treated as a `closed-box' in these theories, but rather
mass is added on long timescales, both as gas and as dissipationless
material, namely the stars and dark matter of the halo
substructure/dwarf galaxy satellites.  The recent discovery of a
satellite galaxy apparently in the process of being accreted by the
Milky Way (Ibata, Gilmore and Irwin, 1994) is a graphic illustration of
this process.

Since the pioneering work of Toomre (e.g., 1977) much effort has gone
into study of the effects of both stellar and gaseous mergers upon
galaxy morphology.   Provided all the orbital energy of an accreted
satellite, mass $M_{sat}$, is available to increase the random
energies of the stars in a thin disk of mass $M_{disk}$, then after a
merger the thin disk will be heated by an amount (Ostriker, 1990)
$$\Delta v_{random}^2 = v_{orbit}^2 M_{sat}/M_{disk}.$$ Of course the
internal degrees of freedom of the satellite could also be excited and
any gas present could, after being heated, cool by radiation, so this
is a definite upper limit to the heating of the disk.
 Note although there
is indeed an age--velocity dispersion relationship for stars in the
thin disk, the  vertical velocity dispersion
 saturates at $\sim 20 $ km/s for stars older than a few Gyr (\eg
Freeman, 1991). This may be understood if the source of the heating
 is confined to the thin disk itself, such as
encounters with giant molecular clouds (\eg reviewed by Lacey, 1991);
some other mechanism is required for the thick disk, which has $\sigma_Z \sim
45$ km/s.

The thick disk of the Milky Way Galaxy at least
morphologically could be   a
remnant of the last significant (in terms of mass ratio)
minor-merger event in the Galaxy's history
 (see e.g. Gilmore,
Wyse and Kuijken, 1989).
The most detailed N-body simulation of this process
 published to date (Quinn, Hernquist and Fullager, 1993) follows
the accretion, by a disk galaxy, of a satellite galaxy with 10\% of
the mass of the disk.  There are many parameters that need to
specified, relating to the initial orbit and to the internal density
profile and kinematics of the satellite, and of the disk. The generic
evolution is that the satellite's vertical motion is damped rather
quickly, with a slower radial decay to the central regions (displayed
very clearly in their Figure 11).  Both the radial and vertical
structures of the disk are altered; the vertical heating increases the
disk scale-height by about a factor of two.  However, the orbital
energy is indeed deposited both in the disk particles and in the
internal degrees of freedom of the satellite; the final galaxy has a
thick disk which consists of both heated thin-disk stars and
shredded-satellite stars.  The mix of these obviously depends on the
(many) model parameters, both orbital and internal to the galaxies.
One may expect the chemical evolution of the disk and of the satellite
to have been different; this should be observable in the chemical
abundances of thick-disk stars.

\section{Chemical Evolution -- The Importance of Element Ratios
and of Scatter}

Chemical evolution models should incorporate the inherent spatial and
temporal inhomogeneity and gas flows of realistic galaxy formation
models -- but avoid getting lost in parameter space.  Elemental
abundances provide a means to break the degeneracy found when simply
overall `metallicity' is followed -- the lack of uniqueness of model
fits to observations of `metallicity' is well documented (e.g. Tosi,
1988). And one should remember that the fact that a model `fits' a
particular observation is not a proof of its appropriateness or its
correctness, but rather that one should move to the next level of
prediction/comparison with observation.  The overwhelming majority of
published models for the chemical evolution of the Galaxy are
concerned only with fitting the mean trend with time, assuming a
one-to-one correlation between time and metallicity.  This is due in
part to the large observational uncertainties of earlier datasets (see
Tinsley, 1975, who never-the-less investigated the possible influence of
a spread in metallicity on the G-dwarf problem).  However, as
discussed by Edvardsson \etal\  (1993) and by Nissen (this
volume), real intrinsic scatter has now been unambiguously
detected in the age-metallicity relation for solar neighbourhood samples of F/G
stars, with scatter of
a few Gyr at given metallicity, or a few tenths of a dex at given
age.  Indeed, one could plausibly argue that the scatter is as large
as any mean trend.  This scatter has to be taken into account in
models  and provides more stringent
constraints on theories.  Star formation should be modelled as a local
process, within molecular cloud complexes, forming stellar
associations with internal enrichment prior to disruption
(\cf\ Tinsley, 1976)
and possible re-formation (Gilmore and Wyse, in prep.).

Element ratios, by contrast, show little intrinsic scatter
(e.g.~Edvardsson \etal, 1993; Nissen, this volume).  Element ratios are
very important for understanding galactic evolution, in part since
different elements come from different sources, and are ejected into
the interstellar medium on different timescales (Tinsley, 1979).  For
example, oxygen is predominantly synthesized in short-lived, massive
stars that explode in Type II supernovae, while iron has an additional
important source in Type Ia supernovae.  The best, but
by no means generally accepted, current model for these latter supernovae
invokes the eventual merging, driven by gravitational radiation, of a
pair of CO white dwarfs, with total mass greater than the Chandrasekhar
mass (Iben and Tutukov, 1987). Thus the explosion timescale is set not
by intrinsic stellar parameters, but rather by orbital parameters, and
can be a Hubble time (note that this is a problem for advocates of
white dwarfs as the dark matter in the haloes of galaxies, unless one
suppresses the binary fraction of the progenitor stars by several
orders of magnitude below that of even the old, metal-poor stars in
our Galaxy; Smecker and Wyse, 1991).  Detailed models suggest that of
order half of the possible progenitors (white dwarf binaries of
suitable orbits and masses) from a single burst of star formation
explode in $\sim 10^9 {\rm yr}$ (\eg\ Smecker-Hane and Wyse, 1992).
The ratio of oxygen-to-iron then is determined in large part by the
ratio of present-to-past star formation rates.

Current understanding of supernova rates and nucleosynthesis implies
that for a solar neighbourhood IMF, Type Ia and Type II supernovae
each provide a yield
of somewhat below solar iron (\cf\ Rana, 1991).  Thus a stellar population
whose mean iron abundance is significantly above solar
either self-enriched sufficiently slowly that Type Ia supernovae
contributed, or had an IMF rather biased towards massive stars.  This
latter possibility will be detectable in the value of the element
ratios.  The dependence on IMF arises since the mass of oxygen ejected
in a Type II supernova is a steeply increasing function of progenitor
mass, while iron, if anything, decreases with progenitor mass (see
Wyse and Gilmore, 1992). The uncertainties in supernova nucleosynthesis
preclude a robust prediction of the [O/Fe] ratio typical from a
generation of massive stars, but {\it offsets\/} as one varies the
massive-star IMF away from that of the solar-neighbourhod can be made
with some confidence.  The simplest analysis is done for a power-law
IMF. An IMF biased towards massive stars by a change of slope of order
unity in the power-law index will result in an enhanced [O/Fe] of
about 0.3 above that value -- itself about 0.3, or a factor of two
above the solar ratio -- for a solar neighbourhood massive-star IMF.

To illustrate the utility of elemental abundances, consider three
independent star-forming regions, each with its own star formation
history and IMF. The pattern of [O/Fe] {\it versus\/} [Fe/H] expected
in stars formed therein is shown schematically in Figure 1.
Generically, one predicts a plateau at early stages, when Type II
supernovae dominate the chemical enrichment, followed by a downturn to
lower values of [O/Fe] after the elapsed time is long enough for Type
I supernovae to have occurred and their iron to be incorporated in the
interstellar medium, and thus into subsequent generations of stars.
It is clearly important to determine the iron abundance at which the
downturn occurs, since this allows one to assign a (relative) age to
this metallicity (\cf\ Matteucci and Greggio, 1986; Truran, 1987;
Wyse and Gilmore, 1988;
Pagel, 1989a, 1994).

\begin{figure}
\vspace{8cm}
\caption{Schematic description  of the
[O/Fe] element ratio evolution as a function of [Fe/H] for
three independent star-forming regions of
differing massive-star IMFs and star formation rates.}
\end{figure}

The dashed line in Figure 1 indicates the evolution of a region with
an IMF dominated by massive stars, compared to the other two, which
have a massive star IMF similar to the population represented now by
the halo subdwarfs.  The biased IMF results in a higher enhancement of
[O/Fe] for the mean Type II supernova, while the higher
nucleosynthetic yield of this region means that, for a given
star-formation rate, a higher [Fe/H] can be reached prior to the onset of
Type Ia supernovae, which occurs at a fixed time after the initiation
of star formation in that region. Of
the remaining two regions, the solid line indicates the evolution with
a higher star-formation rate at early times, and so has more
enrichment through successive generations of massive stars prior to
the onset of chemically-detectable Type Ia supernovae.
An important feature of chemical enrichment by Type I supernovae is
that it can occur {\it without\/} accompanying star formation, since
in most models, the time lag between the formation of the progenitor
stars and the explosion is set by binary orbital parameters and can be
very long.  Thus a system where the star formation occurs in bursts,
separated by periods of little or no star formation, can show very
different patterns of element ratios compared to monotonic star
formation (e.g. application to the Magellanic Clouds by Gilmore and
Wyse, 1991).  Such a possibility is indicated schematically by the
upturn in the dot-dashed line in the Figure, which would happen if the
star-formation rate abruptly increased significantly, providing many
more Type II supernovae per Type I supernova.  A true hiatus in star
formation would be detectable in element ratios by a gap in
oxygen-to-iron versus iron, since iron enrichment from Type I
supernovae would continue during the hiatus and subsequent stars to
form will have incorporated this `extra' iron, without accompanying
oxygen.

\section {Application to the Bulge}

The pattern of [O/Fe] {\it versus\/} [Fe/H] for stars observed at the
solar neighbourhood is similar to the solid line in
Figure 1 (e.g. Wheeler, Sneden and Truran, 1989). The two notable
features of the data most relevant for this discussion are firstly
that there is little detectable scatter about the mean trend, and
secondly that the element ratios are correlated with kinematics --
 stars with halo kinematics have enhanced [O/Fe]
(e.g. Barbuy and Erdelyi-Mendes, 1989, their Table 4).
The hierarchical-merging
picture for the formation of the Galaxy outlined above might lead one
to expect a random mix of all the patterns indicated in Figure 1.
That this does {\it not\/} occur indicates that the massive star IMF
is constant in time and place (e.g. Gilmore and Wyse, 1993; Nissen
\etal, 1994) and further that chemical evolution in the halo
substructure is truncated prior to the onset of Type I supernovae,
i.e. that star formation in individual halo regions is short-lived
(e.g Wyse and Gilmore, 1988), perhaps due to Type II
supernova-driven winds as discussed in section 2 above.
This does {\it not\/} mean that the global halo star formation phase
was short-lived, since the onset of Type I supernovae occurs relative
to the initiation of star formation at any particular location, and
different parts of halo substructure can initiate star formation at
different absolute times (again, as discussed in section 2).
However, the iron abundance at which star
formation is truncated in halo substructure may well vary, reflecting
intrinsic variation in star-formation rates; the absence of stars
with element ratios that are consistent with forming in a manner
indicated by the dot-dashed line in Figure 1 (for example) means that
the gas leftover from halo substructure did not form stars which have
been detected at the solar neighbourhood.  Proposing that this gas
flows towards the Galactic center, to form the bulge, provides an
explanation, and is consistent with angular momentum arguments.
Removal of gas from the halo has been advocated for years as a means
of understanding the low mean enrichment of the halo, assuming an
invariant IMF (as the element ratio data now require); gas loss of
around ten times the mass remaining in  the stellar halo is
suggested (Hartwick, 1976), consistent with dynamical analyses of the
bulge(/bar).  Thus the inhomogeneous model for halo formation that is
inherent in CDM-type models requires incorporation of the monolithic
collapse and radial infall of Eggen, Lynden-Bell and Sandage (1962).
Indeed, specific calculation of element ratios in a model where the
halo pre-enriches the disk and one has `contemporaneous evolution
of two coupled zones' (e.g.
Ferrini \etal, 1992) predicts
unobserved scatter in the element ratios.

The mean iron abundance of the bulge population was previously
estimated to be well above solar metallicity (e.g. Rich, 1988).
However, the most recent, high dispersion (echelle)
spectroscopy of bulge K giants in Baade's Window (McWilliam and Rich,
1994) has discovered a systematic overestimate in iron abundance in
the earlier, lower-resolution, studies, probably due to  use of
combined Mg and Fe features, implicitly assuming solar element ratios,
and to molecular line blanketing.  The overestimate is most severe for
the highest chemical abundances, ranging from only about 0.1 dex at
[Fe/H]$ = -1$dex to $\sim +0.5$dex at [Fe/H]$=+0.5$dex.  (The
ramifications of this for the utility of the bulge as a template for
external elliptical galaxies have yet to be fully investigated.)  The
new results, based on a transformation derived from  a
sample of ten stars chosen from Rich's sample to
cover the range of iron abundance,
give a mean iron abundance of $-0.3$dex.

Element ratios will also be important discriminants among models of
star formation in the bulge, constraining the accumulation time for
material at the center and the star formation timescale.  If the
proto-bulge material accumulated, and formed stars, sufficiently
rapidly that only Type II supernovae are available for chemical
enrichment, then the element ratios reflect the massive star IMF.
Unfortunately the rather sparse elemental abundance data, for only 10
stars (McWilliam and Rich, 1994), are rather confusing, with Mg and Ti
showing enhancements relative to iron indicative of Type II enrichment
(with a solar neighbourhood IMF) over the entire range of iron
abundances, Si and Ca, in contrast, having the solar ratio relative to
iron for stars more iron-rich than $-0.5$ dex, and [O/Fe] being
consistent with solar for all stars.  Clearly more data are required.

If the bulge formed `rapidly', then after star formation has ceased
iron will eventually be
synthesised in Type~I supernova and ejected into the bulge, releasing
about the same amount of iron as ejected by Type~II supernovae, for a
reasonable range of IMFs and models of Type~I supernovae (the Type~I
supernovae from the halo background population that is also present
are merely perturbations to the present model, due to the much larger
mass, $\sim \times 10$, of the bulge). This gas must be accounted for
in the model.  Such iron-rich gas, which is not incorporated in the
main bulge population, is produced in any model for the bulge which
has sufficiently rapid star formation that it is essentially complete
prior to Type~I explosions.  Should this gas sustain a low level of
star formation, then one may expect to find extremely iron-rich stars
which have very low oxygen abundances (\cf\ Gilmore and Wyse, 1991).
The relatively slow, and uncorrelated, rate at which Type I supernovae
explode contrasts with Type II supernovae, and it is unlikely that the
iron-rich gas will be ejected as a wind, but depending on its density,
it may be maintained at a high temperature; the recent detection of
X-ray emitting gas at the center may be related (Yamauchi \etal, 1991;
Blitz \etal, 1993).

More direct estimation of the age of the dominant population in the
bulge can be attempted by analysis of colour-magnitude diagrams.
Fairly model-independently, one may expect the first star formation in
the Galaxy to have occurred in the densest regions, which have the
shortest dynamical times.  The high spatial resolution of HST allows
fitting of the main-sequence turn-off, relatively unaffected by
crowding.  Analysis of both the luminosity function and turnoff of
unevolved bulge stars in Baade's window (Holtzman \etal, 1993)
suggests a typical age of only 5 -- 10 Gyr for a mean metallicity of
around half the solar abundance, as the most recent observations
imply; an exclusively old bulge is only possible with unlikely
combinations of reddening, distance and photometric uncertainties.
Extremely well-populated colour-magnitude diagrams of the evolved
stars in the bulge have become available as a side-benefit of searches
for gravitational micro-lensing events and allow study of
shorter-lived (and hence rarer) evolutionary phases.  In particular
the OGLE collaboration find a substantial population of
core-helium-burning `red clump' stars which they attribute to a large
intermediate-age population, significantly younger than 10Gyr
(Paczsynski \etal, 1994).  This is also consistent with the inferences
from analysis of the periods of variable stars in the bulge
(e.g. Harmon and Gilmore, 1988).  However, Renzini (1994) points out
that the lifetime of the `clump' is also sensitive to helium content,
and the relative populations of `clump' and red giant stars may be
understandable with an old age but enhanced helium (which obviously
depends on the parameter $\Delta Y/ \Delta Z$).  The presence of
significantly older stars has been inferred from the horizontal branch
morphology, in particular the relative number of RR Lyrae stars, by
Lee (1992 -- but see also Wyse, 1993).  At present the age {\it
distribution\/} in the bulge remains rather uncertain.

The presence or otherwise of chemical (and elemental) abundance
gradients in the bulge contains clues to bulge formation.  A lack of a
gradient argues against dissipational formation.  The McWilliam and
Rich data pertain only to Baade's window, of order  500pc from  the
Galactic Center. Ibata and Gilmore (1994) have achieved `long-slit
spectroscopy' of the bulge analagous to that done for external
galaxies through two series of multi-fibre fields, one parallel to the
minor axis, but at $R \sim 3$kpc, and one parallel to the major axis,
at $Z \sim 3$kpc.  Their total sample is around 2,000 K giants.  This
shows no evidence for gradients, and has a mean metallicity
(albeit calibrated using  a magnesium feature) of $-0.3$dex,
with a distribution that is remarkably similar to that derived for
Baade's window by McWilliam and Rich.  Further, the K-giants in the
inner disk also have a mean metallicity of $-0.3$ dex (Lewis and
Freeman, 1989).  Thus the overall chemical enrichment of the bulge is
consistent with a dissipationless formation of the bulge from the
inner
disk.  Such a formation mechanism for the bulge has indeed been
proposed -- an out-of-plane bending instability in a bar-unstable disk
(Combes \etal, 1990;  Raha \etal, 1991).   The morphology of the
central regions of the galaxy after the bending instability has
fattened the bar looks remarkably like the COBE images of our Galaxy,
even with a distinct `peanut' minimum along the minor axis.  However,
the latest analysis of the DIRBE data (Weiland \etal, 1994) concludes
that the peanut morphology is an artefact of varying extinction.
Also, the scaleheight of the COBE bulge is around 300pc, which is equal
to that of the old disk in the solar neighbourhood, despite
significant fattening of the bulge relative to the disk being an
important feature of the instability.  Now, the disk as seen by DIRBE
and in the ground-based IR observations is significantly thinner than
the local old disk (Kent \etal, 1991), but this may be due to the fact
that the {\it young\/} disk is preferentially detected -- observations
of edge-on external galaxies consistently show the (old) disk
scale-height to be independent of radius (e.g. van der Kruit and
Searle, 1981; Shaw and Gilmore, 1990).  Detailed analyses to understand
which stellar populations are studied in the IR are needed.

The kinematics of bulge stars -- as evidenced in the angular
momentum distribution -- suggest a closer connection to the halo
than to the present disk; detailed predictions for the kinematics of bulge
stars in the bending instability model should be made.
It may be that the halo ejecta forms the inner disk and still
satisfies angular-momentum and element-ratio constraints.
It is also noteworthy
that there is as yet no compelling evidence from stellar kinematics
that the dominant gravitating mass of the bulge region is triaxial; the oblate
rotator model (e.g. Kent, 1992) provides an acceptable fit (e.g. Ibata and
Gilmore, 1994).  Perhaps the firmest indication for triaxiality
is provided by the analysis of Zhao, Spergel and Rich
 (1994) of bulge stars with both
proper motions and radial velocities, but this is a small sample with large
error bars.
The implications for chemical evolution of the inferred triaxial
nature of the central regions of the Galaxy (e.g. Blitz and Spergel,
1991; Binney \etal, 1991) have not been explored.  One obvious effect
is the enhanced inflow of gas expected due to the gravitational
torques of the bar, perhaps of amplitude 0.01 -- 0.1 M$_\odot$/yr
(Gerhard and Binney, 1993).

\section {Application to the Disk}

The sample of disk F/G stars of Edvardsson \etal\ (1993;
discussed by Nissen, this volume) has provided an unprecedented
opportunity to study the pattern of element ratios for stars for
which independent age estimates are  available.  This allows the
star-formation history to be analysed both
in terms of  relative timescales -- relative to the onset of star formation
in a particular local star-forming region -- and of an absolute
timescale in the global scheme of the formation of the Galaxy.
Referring back to Figure 1, stars which formed with a higher
star formation rate will sustain the plateau of enhanced [O/Fe]
to higher [Fe/H]; this is seen in the Edvardsson \etal\ sample
once stars which are believed (from their kinematics plus a model
Galactic potential) to have formed in the inner Galaxy are isolated.  Indeed,
the data are in good  agreement with models that predict the
disk formed stars more rapidly in its inner regions, with a
smooth variation of star formation rate with radius (e.g. Wyse
and Silk, 1989).

The observed lack of scatter in the trend of element ratios, once
a sub-sample formed at a given radius has been isolated, places
severe constraints on the  variation of the star formation
rate and of the IMF (e.g.~Wyse and Gilmore, 1988).  A
monotonically-declining star formation rate is favoured, with a
constant IMF.

The Edvardsson \etal\ sample was defined at the solar neighbourhood,
 and thus is biased kinematically, so that trends of element ratios
 with kinematics cannot be easily investigated.  However, the element
 ratios of thick-disk stars appear to match smoothly onto the trend
 set by the halo stars, with little or no scatter,
and as argued above this can only arise within
 the context of inhomogeneous halo models if the IMF is
invariant, and there is no chemical
 connection between  thick disk and halo.
 This raises the as-yet-unanswered question of what pre-enriches the
 thick disk, if there is indeed no significant `metal-poor thick
 disk', as  implied by the current observations which characterise
 the thick disk as pre-dominantly stars with iron abundance of
 $-0.6$dex (see Gilmore, Wyse and Jones, 1995; Ryan and Lambert, 1995).

The lack of scatter in element ratios contrasts with the large scatter
in the age-metallicity relationship of this sample;
intermediate age stars are seen with essentially the entire range of
iron abundance in the sample, from one tenth of solar to above solar.
The trend of decreasing age with increasing metallicity is due to
essentially only the youngest and oldest stars.  It should be noted
that stars of approximately half the solar metallicity existed 10Gyr
ago, which is relevant to the interpretation of qso absorption-line
systems.  Also the age determinations of individual thick-disk stars
agree with the more indirect inferences from the metallicity and
turn-off B$-$V colour of the thick disk, $\sim 15$Gyr (Wyse and
Gilmore, 1988). This relatively old age  contrasts with that
inferred for the disk from analysis of the white dwarf luminosity
function (e.g. Winget \etal, 1987).  Taking both results at face
value suggests that there may have been a hiatus in star formation
between the earliest disk (now the thick disk) and the main body of
the disk.  However, an age gap is not seen in the Edvardsson \etal\
dataset, nor is there any discontinuity in element ratios, as
would be expected for a substantial hiatus (see section 3 above).

The scatter in the age-metallicity relationship in the sample of
Edvardsson \etal\ remains even if one restricts the sample to stars
one believes formed in a narrow radial range;   This suggests that one cannot
appeal to a
mean abundance gradient (as inherent in models where the inner regions
evolve faster, as inferred from the element ratio data) combined with
epicyclic orbits to mix stars (but see Fuchs and Weilen's
contributed paper this volume).  Rather, star formation and chemical
evolution are inherently inhomogeneous.

The scatter seen in the age-metallicity relationship
plausibly reflects the inhomogeneity of star formation within
giant molecular cloud complexes in the disk, provided that star
formation regions self-enrich, and GMC complexes
dissociate and reform with incomplete mixing of chemical
elements.  Alternatively, one can appeal to infall of
metal-poor gas that forms star prior to complete mixing (White
and Audouze, 1980; Edvardsson \etal, 1993).
 Both these scenarios  require that different elements mix
the same way, despite their different production sites and
timescales.
Observational input comes from the detailed elemental abundances
of main sequence B stars in the four main age subgroups
in the Orion association   (Cunha and
Lambert, 1992). The age difference is $\sim 10^7$yr across the
association, and  the youngest
subgroup is enhanced in oxygen by 0.2dex relative to the older
subgroups.   Although this spread in oxygen seems high, should
it be due to enrichment from a typical Type II supernova in the
older subgroups, and the iron from that supernova mixed with the
oxygen, this produces a trend that is
within the limits of scatter set by  the element
data of Edvardsson \etal\ at the iron abundance of Orion.

At present no model simultaneously and self-consistently predicts a
high amplitude of scatter in the age-metallicity relationship and no
detectable scatter in the element ratios.  Classical tests of chemical
evolution in the thin disk, such as the G-dwarf metallicity
distribution which describes the integrated chemical enrichment
history of the solar neighbourhood, are predicated on the assumption
of a steady increase of metallicity with time, and are difficult to
interpret when the dispersion in metallicity at a fixed time is
approximately equal to the dispersion when one integrates over time
(as pointed out by Tinsley, 1975).  Malinie \etal\ (1993) developed a
model of inhomogeneous chemical evolution by adopting {\it a
priori\/} the amplitude of
scatter seen in the dataset of Edvardsson \etal\ However, they did not
attempt self-consistent modelling and to fit the G-dwarf metallicity
distribution they were forced as usual simply to  assume that the disk
was pre-enriched to 0.3 solar abundance (e.g. Pagel and Patchett, 1975;
Pagel, 1989b).  Gilmore and Wyse (1986) demonstrated that the thick
disk, if sufficiently massive, could provide the required
self-consistent pre-enrichment of the thin disk.  However, that paper
appealed to the halo to pre-enrich the thick disk (see also
Burkert, Truran and Hensler, 1992).  Updated samples of F/G stars, with
improved metallicity determinations and selection criteria, should
provide more detailed testing of solar-neighbourhood chemical
evolution (Gilmore and Wyse, 1995; Bikmaev, this volume).

A possible shortcoming of the Malinie \etal\ treatment of chemical
inhomogeneity is that they assume it to be shortlived and erased
between episodes of star formation. We (Gilmore and Wyse) are
developing models that rather allow inhomogeneities to survive the
dispersal and re-formation of the giant molecular cloud complexes in
which stars form.  This may more accurately reflect the physical
processes that operate (e.g. Elmegreen, 1979), particularly in light of
the realisation (e.g. Wyse and Silk, 1985;
Heiles, 1990; Norman and Ikeuchi, 1989) that the vertical
potential of the disk is sufficiently weak that gas
from the disk is probably re-cycled through ejection into the halo
where it can cool, perhaps  without significant mixing.


\begin{thebibliography}{}

\bibitem[]{}
Barbuy, B. and Erdeyli-Mendes, R., 1989, {\it A. \&\ A.} {\bf 214}, 239

\bibitem[]{}
Binney, J.J., \etal,
1991, {\it M. N. R. A. S.} {\bf 252}, 210

\bibitem[]{}
Blitz, L. and Spergel, D.N., 1991, {\it Ap. J.} {\bf 379}, 631

\bibitem[]{}
Blitz, L. Binney, J., Lo, K.Y., Bally, J. and Ho, P.T.P., 1993,
{\it Nature} {\bf 361}, 417

\bibitem[]{}
Blumenthal, G.R., Faber, S.M., Primack, J.R. and
Rees, M.J., 1984, {\it Nature} {\bf 311}, 517

\bibitem[]{}
Burkert, A., Truran, J. and Hensler, G., 1992, {\it Ap. J.} {\bf 391}, 651

\bibitem[]{}
Carney, B., 1990, {\it Bulges of Galaxies}, eds. B.Jarvis
and D.Terndrup, ESO,  Garching, p.26


\bibitem[]{}
Combes, F., Debbasch, F., Friedl, D. and Pfenniger, D., 1990, {\it A. \&\ A.}
{\bf 233}, 82

\bibitem[]{}
Cunha, K. and Lambert, D., 1992, {\it Ap. J.} {\bf 399}, 586

\bibitem[]{}
Dekel, A. and Silk, J., 1986, {\it Ap. J.} {\bf 303}, 39

\bibitem[]{}
Edvardsson, B., \etal,
 1993, {\it A. \& A.} {\bf 275}, 101

\bibitem[]{}
Eggen, O.J., Lynden-Bell, D. and Sandage, A.R., 1962, {\it Ap. J.}
{\bf 136}, 748

\bibitem[]{}
Elmegreen, B., 1979, {\it Ap. J.} {\bf 231}, 372

\bibitem[]{}
Fall, S.M. and Efstathiou, G.P., 1980, {\it M. N. R. A. S.} {\bf 193}, 189

\bibitem[]{}
Ferrini, F., Matteucci, F., Pardi, C. and  Penco, U., 1992, {\it Ap. J.} {\bf
387}, 138

\bibitem[]{}
Freeman, K.C., 1991, {\it Dynamics of Disc Galaxies}, ed. B.~Sundelius,
G\"oteborgs University, G\"oteborg,  p.15


\bibitem[]{}
Gerhard, O.E. and Binney, J., 1993, {\it Galactic Bulges}, eds.
H.~Dejonghe and H.J.~Habing, Kluwer, Dordrecht, p.275

\bibitem[]{}
Gilmore, G. and Wyse, R.F.G., 1995, submitted to {\it A. J.}

\bibitem[]{}
Gilmore, G. and Wyse, R.F.G.,
1993, {\it First Light in the Universe},
8th IAP Meeting, eds B.~Rocca-Volmerange, M.~Dennefeld,
B.~Guiderdoni and J.~Tran Than Van, p.177

\bibitem[]{}
Gilmore, G. and Wyse, R.F.G., 1991, {\it Ap. J.} {\bf 367}, L55

\bibitem[]{}
Gilmore, G. and Wyse, R.F.G., 1986, {\it Nature} {\bf 322}, 806

\bibitem[]{}
Gilmore, G., Wyse, R.F.G. and Jones, J.B., 1995, {\it A. J.}, in press

\bibitem[]{}
Gilmore, G., Wyse, R.F.G. and Kuijken, K., 1989, {\it Ann. Rev.
Astron. Astroph.} {\bf 27}, 555

\bibitem[]{}
Gunn, J.E., 1987, {\it The Galaxy}, eds. G.~Gilmore and
B.~Carswell, Reidel, Dordrecht, p.413

\bibitem[]{}
Harmon, R.A. and Gilmore, G., 1988, {\it M. N. R. A. S.} {\bf 235}, 1025

\bibitem[]{}
Heiles, C., 1990, {\it Ap. J.} {\bf 354}, 483

\bibitem[]{}
Holtzman, J.A., \etal,
1993, {\it A. J.} {\bf 106}, 1826

\bibitem[]{}
Ibata, R., Gilmore, G., and Irwin, M., 1994, {\it Nature} {\bf 370}, 194

\bibitem[]{}
Hartwick, F.D.A., 1976, {\it Ap. J.} {\bf 209}, 418

\bibitem[]{}
Iben, I. and Tutukov, A.V., 1987, {\it Ap. J.} {\bf 313}, 727



\bibitem[]{}
Kent, S.M., 1992, {\it Ap. J.} {\bf 387}, 181

\bibitem[]{}
Kent, S.M., Dame, T.M. and Fazio, G., 1991, {\it Ap. J.} {\bf 378}, 131

\bibitem[]{}
Kruit, P.C. van der and Searle, L., 1981, {\it A. \&\ A.} {\bf 95}, 105

\bibitem[]{}
Lacey, C.G., 1991, {\it Dynamics of Disc Galaxies},
ed. B.~Sundelius, G\"oteborgs University, G\"oteborg, p.257

\bibitem[]{}
Lacey, C.G. and Cole, S., 1993, {\it M. N. R. A. S.} {\bf 262}, 627

\bibitem[]{}
Lee, Y.-W., 1992, {\it A. J.} {\bf 104}, 1780

\bibitem[]{}
Lewis, J. and Freeman, K.C., 1989, {\it A. J.} {\bf 97}, 139

\bibitem[]{}
Malinie, G., Hartmann, D.H., Clayton, D.D. and Mathews,
G.J., 1993, {\it Ap. J.} {\bf 413}, 633


\bibitem[]{}
Matteucci, F. and    Greggio, L., 1986, {\it A. \&\ A.} {\bf 154}, 279

\bibitem[]{}
McWilliam, A. and Rich, M., 1994, {\it Ap. J. Suppl.} {\bf 91}, 749

\bibitem[]{}
Navarro, J., Frenk, C.S. and White, S.D.M., 1994, {\it M. N. R. A. S.}
{\bf 267}, L1

\bibitem[]{}
Nissen, P.E., Gustafsson, B., Edvardsson, B. and Gilmore, G.,
1994, {\it A. \&\ A.}, in press


\bibitem[]{}
Norman, C.A. and Ikeuchi, S., 1989, {\it Ap. J.} {\bf 345}, 372

\bibitem[]{}
Ostriker, J.P., 1990, {\it Evolution of the Universe of
Galaxies}, ed. R.G.~Kron, A.S.P. Conf.
Series {\bf 10}, A.S.P., San
Francisco, p.25

\bibitem[]{}
Paczynski, B., \etal,
 1994, {\it A. J.} {\bf 107}, 2060

\bibitem[]{}
Pagel, B.E.J., 1994, {\it IAC Winter School on Galaxy Formation
and Evolution}, ed. C.~Munoz-Turon, CUP, Cambridge, in press

\bibitem[]{}
Pagel, B.E.J., 1989a, {\it Rev.~Mex.~Astr.~Astrofis.} {\bf 18}, 153

\bibitem[]{}
Pagel, B.E.J., 1989b,
{\it Evolutionary Phenomena in Galaxies}, eds.
J.~Beckman and B.E.J.~Pagel, CUP, Cambridge, p.201

\bibitem[]{}
Pagel, B.E.J. and Patchett, B.E., 1975, {\it M. N. R. A. S.} {\bf 172}, 13

\bibitem[]{}
Quinn, P.J., Hernquist, L. and Fullager,
D.P., 1993, {\it Ap. J.} {\bf 403}, 74

\bibitem[]{}
Quinn, T. and Binney, J.J., 1992, {\it M. N. R. A. S.} {\bf 255}, 729

\bibitem[]{}
Raha, A., Sellwood, J., James, R. and Kahn, F.D., 1991, {\it Nature}
{\bf 352}, 411

\bibitem[]{}
Rana, N.C., 1991, {\it Ann. Rev. Astron. Astroph.} {\bf 29}, 129

\bibitem[]{}
Renzini, A., 1994, {\it A. \&\ A.} {\bf 285}, L5

\bibitem[]{}
Rich, R.M., 1988, {\it A. J.} {\bf 95}, 828

\bibitem[]{}
Ryan, S.J. and Lambert, D.A., 1995, {\it A. J.}, in press

\bibitem[]{}
Ryden, B.S., 1988, {\it Ap. J.} {\bf 329}, 589


\bibitem[]{}
Searle, L. and Zinn, R., 1978, {\it Ap. J.} {\bf 225}, 357


\bibitem[]{}
Shaw, M.A. and Gilmore, G., 1990, {\it M. N. R. A. S.} {\bf 242}, 59

\bibitem[]{}
Silk, J. and Wyse, R.F.G., 1993, {\it Phys. Rep.} {\bf 231}, 293

\bibitem[]{}
Smecker, T.A. and Wyse, R.F.G., 1991, {\it Ap. J.} {\bf 372}, 448

\bibitem[]{}
Smecker-Hane, T.A. and Wyse, R.F.G., 1992, {\it A. J.} {\bf 103}, 1621


\bibitem[]{}
Tinsley, B.M., 1975, {\it Ap. J.} {\bf 197}, 159

\bibitem[]{}
Tinsley, B.M., 1976, {\it Ap. J.} {\bf 208}, 797

\bibitem[]{}
Tinsley, B.M., 1979, {\it Ap. J.} {\bf 229}, 1046


\bibitem[]{}
Tosi, M., 1988, {\it A. \&\ A.} {\bf 197}, 33


\bibitem[]{}
Toomre, A., 1977, {\it The Evolution of Galaxies and
Stellar Populations}, eds. B.M.~Tinsley and R.B.~Larson, Yale
Univ. Obs, New Haven, p.401



\bibitem[]{}
Truran, J.W., 1987, {\it 13th Texas Symposium on Relativistic
Astrophysics}, ed. M.P. Ulmer, Singapore, World Scientific, p.430

\bibitem[]{}
Weiland, J.L. {\it et al.}, 1994, {\it Ap. J.} {\bf 425}, L81

\bibitem[]{}
Wheeler, J.C., Sneden, C. and Truran, J.W., 1989, {\it Ann. Rev. Astron.
Astroph.} {\bf 27}, 279

\bibitem[]{}
White, S.D.M and Audouze, J., 1983, {\it M. N. R. A. S.} {\bf 203}, 603


\bibitem[]{}
Winget, D.E. {\it et al.}, 1987, {\it Ap. J.} {\bf 315}, L77

\bibitem[]{}
Wyse, R.F.G., 1993, {\it Nature} {\bf 361}, 204

\bibitem[]{}
Wyse, R.F.G. and Gilmore, G., 1988, {\it A. J.} {\bf 95}, 1404

\bibitem[]{}
Wyse, R.F.G. and Gilmore, G., 1992, {\it A. J.} {\bf 104}, 144

\bibitem[]{}
Wyse, R.F.G. and Silk, J., 1985, {\it Ap. J.} {\bf 296}, L1

\bibitem[]{}
Wyse, R.F.G. and Silk, J., 1989, {\it Ap. J.} {\bf 339}, 700

\bibitem[]{}
Yamauchi, Y. {\it et al.}, 1991, {\it Ap. J.} {\bf 365}, 532


\bibitem[]{}
Zhao, H., Spergel, D.N. and Rich, R.M., 1994, preprint

\bibitem[]{}
Zurek, W., Quinn, P. and Salmon, J., 1988, {\it Ap. J.} {\bf 330}, 519

\end{thebibliography}
\end{document}